\documentclass{ws-procs9x6}

\begin{document}

\title{Nucleon spin structure and its connections to various sum rules} 

\author{J. SOFFER}

\address{Centre de Physique Th\'eorique, \\
UMR 6207 \footnote{\uppercase{UMR} 6207 is \uppercase{U}nit\'e \uppercase{M}ixte
de \uppercase{R}echerche du \uppercase{CNRS} and of \uppercase{U}niversit\'es
\uppercase{A}ix-\uppercase{M}arseille \uppercase{I}
and \uppercase{A}ix-\uppercase{M}arseille \uppercase{II}, and of 
\uppercase{U}niversit\'e 
du \uppercase{S}ud \uppercase{T}oulon-\uppercase{V}ar, \uppercase{L}aboratoire
affili\'e \`a la \uppercase{FRUMAM}} CNRS-Luminy Case 907,\\
13288 Marseille Cedex 9, France \\ 
E-mail: soffer@cpt.univ-mrs.fr}

\maketitle

\abstracts{
Our knowledge on the nucleon spin structure has greatly improved over
the last twenty years or so, but still many fundamental questions remain 
unsolved. I will try to review some of the puzzling aspects of the origin 
of the nucleon spin. I will emphasize the connection with several sum 
rules and, when using this tool, the relevance of some kinematic regions
for testing them in the QCD dynamics framework.}

\section{Introduction}

The reply to the question " {\it why do we need spin in high energy particle physics?}~",
 on very general grounds, is twofold: first we want to learn about
hadron structure and second we want to test perturbative QCD in the spin sector.
More specifically concerning the nucleon structure we need
to determine the unpolarized parton
distributions $ f_N(x,Q^2)$ ($N=p,n$ for proton and neutron), where $f$ stands for
quarks ($u,d,s,..$), antiquarks ($\bar {u},\bar {d},\bar {s},..$) or gluons.
We also need to know the corresponding helicity distributions $ \Delta f_N(x,Q^2)$
and transversity distributions. All these distributions are functions of the scaling
variable $x$ and the $Q^2$ dependence, so the scaling violations predicted by QCD,
must be compared to experimental results. The data allow to extract various structure
functions, unpolarized $F_2^{p,n},F_3,..$ and polarized  $g_1^{p,n}$, 
from deep inelastic scattering (DIS), which are expressed in terms of the parton 
distributions. The relevant experiments are currently performed at CERN (Compass), DESY
(Hermes), JeffLab and SLAC, but one should also mention the new RHIC spin programme 
at the {\it polarized} $pp$ collider at BNL \cite{BSSV}, which will go beyond testing 
the QCD scaling violations of the polarized parton distributions.
  
\section{ Many sum rules on structure functions}

There exists a number of sum rules for unpolarized and polarized
structure functions, some of which are rigorous results and other which rely
on more or less well justified assumptions. Let us first consider
the charged current structure functions in neutrino DIS, 
for which we have two rigourous results, namely the 
{\it Adler sum rule} \cite{ADL} (ASR)
\begin{equation}\label{1}
\int^{1}_{0}\frac{dx}{2x} \left[F^{\bar{\nu}p}_2(x) - F^{\nu
p}_2(x)\right] =
N_u-N_d=1\,
\end{equation}
because
\begin{equation}\label{2}
N_u=\int^{1}_{0}dx\ u_{val}(x)=2\quad\hbox{and} \quad
N_d=\int^{1}_{0}dx\ d_{val}(x)=1\ ,
\end{equation}
and the {\it Gross-Llewellyn Smith sum rule} \cite{GRO} (GLSSR)
\begin{equation}\label{3}
I_{GLS}\!\!=\!\!\int^{1}_{0}\frac{dx}{2x} \left[xF^{\nu p}_3(x) + x
F^{\bar{\nu} p}_3(x)\right] \!\!=\!\!
N_u+N_d\!= \!3 .
\end{equation}
The ASR is exact and receives no QCD corrections, but its experimental
verification is at a very low level of accuracy \cite{ALLA}. The GLSSR gets
a negative QCD correction and the CCFR data \cite{QUI} gives
$I_{GLS}=2.55\pm 0.06\pm 0.1$ at $Q^2=3\mbox{GeV}^2$, in fair agreement with
the theoretical prediction.

Next, let us consider the
unpolarized electromagnetic structure functions for proton
and neutron $F^{p,n}_2(x)$, for which we have 
the {\it Gottfried sum rule}\cite{GOT} (GSR). If one assumes
an SU(2) symmetric sea, {\it i.e.} $\bar u(x)=\bar d(x)$, one can
easily show, using Eq.(\ref{2}), that
\begin{equation}\label{4}
I_G=\int^{1}_{0}\frac{dx}{x} \left[F^p_2(x) - F^n_2(x)\right] =
\frac{1}{3}\ .
\end{equation}

In fact the NMC experiment\cite{ARN} has observed a large defect of the
GSR, since their measurement gives at 
$Q^2=4 \mbox{GeV}^2$, $I_G=0.235\pm 0.026$. 
This flavor symmetry breaking, more precisely
$\bar d>\bar u$, is a consequence of the Pauli exclusion principle
which favors $d\bar d$ pairs with respect to $u\bar u$ pairs, since the
proton contains two $u$ quarks and only one $d$ quark.

If we turn to polarized structure functions, there is first a
fundamental result called the {\it Bjorken sum rule}\cite{BJOR} (BSR).
It was derived about thirty years ago in the framework of quark current
algebra and it relates the first moment of the difference between
$g^p_1(x)$ for the proton and $g^n_1(x)$ for the neutron and the
neutron $\beta$-decay axial coupling
\begin{equation}\label{5}
\int^{1}_{0}dx\left[g^p_1(x) - g^n_1(x)\right] =\frac{1}{6} g_A/g_V\ ,
\end{equation}
where $g_A/g_V = 1.2573\pm 0.0028$ is very accurately known. The BSR
gets also a negative QCD correction and we will come back later to the
test of this firm prediction of QCD. One can also derive sum rules for
$g^p_1$ and $g^n_1$ separately. 
These are the {\it Ellis-Jaffe sum rules}\cite{ELL} (EJSR) which read
\begin{eqnarray}
&&\Gamma_1^p= \int^{1}_{0}dx g^p_1(x)=\frac{1}{18}
(9F-D+6\Delta s)\quad\hbox{and}
\nonumber \\
&&\Gamma_1^n=\int^{1}_{0}dx g^n_1(x)=\frac{1}{18} (6F-4D+6\Delta s)\ ,
\label{6}
\end{eqnarray}
where $F=0.459 \pm 0.008$ and $D=0.798 \pm 0.008$ are the $\beta$-decay axial 
coupling constants of the
baryon octet and $\Delta s= \int^{1}_{0}\Delta s(x)dx$ is the total
polarization of the proton carried by the strange quarks. One recovers
Eq.~(\ref{5}) by taking the difference because $F+D= g_A/g_V$. 
In their original work, Ellis and Jaffe made the
critical assumption $\Delta s=0$, which allows to make definite
predictions for $\Gamma^p_1$ and $\Gamma^n_1$.  One gets $\Gamma^p_1=0.19$,
in strong disagreement with the sixteen years old EMC data \cite{EMC} 
$\Gamma^p_1=0.112 \pm 0.009 \pm 0.19$. This large defect, which still
remains, was first attributed
to a large $\Delta s$, but this naive interpretation has been ruled 
out since then, on experimental grounds. One should keep in mind that
to test all the above sum rules, one requieres an accurate determination of the
corresponding structure functions in the full kinematic range $0\leq x \leq 1$,
which is never achieved in current experiments. Significant progress will have to
be achieved, to reduce the missing region, where some dangerous extrapolations are
needed.

Concerning the other polarized structure function $g_2(x)$, which is related to
transverse polarization, but has no simple interpretation 
in the parton model, it is possible to derive a superconvergence relation by
considering the asymptotic behavior of a particular virtual Compton
helicity amplitude. This leads to the 
{\it Burkhardt-Cottingham sum rule}\cite{BUR} (BCSR)
\begin{equation}\label{7}
\int^{1}_{0}dxg^p_2(x)= 0 \quad\hbox{and}  \quad \int^{1}_{0}dxg^n_2(x)=
0\ ,
\end{equation}
for proton and neutron and from this result, it has been naively argued
that $g_2(x)$ vanishes identically. In fact, alternatively a simple 
relation between $g_1$ and $g_2$ can be 
expected, namely $g_1(x) +g_2(x) = 0$. 
However $g_2(x)$ is more complicated than
that \cite{JI} and only part of it (its twist-$2$ contribution) is entirely
related to $g_1(x)$, by means of the {\it Wandzura-Wilczek sum
rule}\cite{WAN} (WWSR) which reads for $J\geq 1$
\begin{equation}\label{8}
\int^{1}_{0}dx x^{J-1}\left[\frac{J-1}{J} g_1(x) + g^{WW}_2(x)\right] =
0\ .
\end{equation}
Clearly for $J=1$ one recovers the BCSR Eq.~(\ref{7}) and for $J=2$ one
has
\begin{equation}\label{9}
g^{WW}_2(x)= - g_1(x) + ~\int^{1}_{x} g_1(y) \frac{dy}{y} \ .
\end{equation}

Finally, using the spin-dependent photoabsorption cross sections $\sigma_{1/2(3/2)}(\nu)$, 
{\it Gerasimov-Drell-Hearn}\cite{GERA} (GDH) 
have derived the following celebrated sum rule valid for real photons
\begin{equation}\label{eq:gdh}
I_{GDH}= \int_{\nu_{thr}}^\infty [\sigma_{1/2}(\nu)-\sigma_{3/2}(\nu)] \frac{d\nu}{\nu} =
- \frac{2\pi^2\alpha}{M^2}\kappa^2 \ ,
\end{equation}
where $\nu$ is the photon energy in the target rest frame,
$\nu_{thr}$ is the pion production threshold, $\kappa$ the anomalous nucleon 
magnetic moment, $M$ the nucleon mass and $\alpha$ the fine structure coupling constant.
However the GDH integral can be generalized to the case of absorption of
polarized transverse virtual photons with $Q^2$
\begin{equation}\label{eq:ggdh}
I_{GDH}(Q^2)=\int_{\nu_{thr}}^\infty[\sigma_{1/2}(\nu,Q^2)-\sigma_{3/2}(\nu,Q^2)]\frac{d\nu}{\nu}.
\end{equation}
One can show that, within a good approximation, one has in the scaling limit
\begin{equation}\label{eq:GDH}
I_{GDH}(Q^2)= \frac{16 \pi^2 \alpha}{Q^2} \Gamma_1 \, ,
\end{equation}
where $\Gamma_1= \int^{1}_{0}dx g_1(x)$, so the GDH sum rule is connected to polarized DIS.
The $Q^2$-dependence of the generalized GDH sum rule has been measured very accurately
recently, with different targets, and we will come back in more details to some theoretical 
understanding of these data. 
  
\section{The statistical approach for polarized parton distributions}

DIS of leptons on hadrons has been extensively
studied, over the last twenty years or so, both theoretically and experimentally, to
extract the polarized parton distributions of the nucleon.
As it is well known, the unpolarized light quarks ($u,d$) distributions are fairly 
well determined. Moreover, the data exhibit a clear evidence for a
flavor-asymmetric light sea, {\it i.e.} $\bar d > \bar u$ as mentioned above,  
and large uncertainties still persist for the gluon ($G$) and the heavy quarks 
($s,c$) distributions. The corresponding polarized gluon and $s$ quark
distributions ($\Delta G, \Delta s$) are badly constrained and we just begin to
uncover a flavor asymmetry, for the corresponding polarized light sea, namely  
$\Delta \bar u \neq \Delta \bar d$. 
Whereas the signs of the polarized light quarks distributions are 
essentially well established, $\Delta u > 0$ and $\Delta d < 0$, 
this is not the case for $\Delta \bar u $ and $\Delta \bar d $.
Here we briefly recall how 
we construct a complete set of polarized parton
(all flavor quarks, antiquarks and gluon) distributions. 
Our motivation is to use the statistical approach \cite{BBS,BBS1} to
build up : $q_i$, $\Delta q_i$, $\bar q_i$, $\Delta \bar q_i$, $G$ and $\Delta
G$, in terms of a very small number of free parameters. 
A flavor separation for the unpolarized and polarized light sea
is automatically achieved in a way dictated by our approach. 

The existence of the correlation, broader shape higher first moment, suggested
by the Pauli principle, has inspired the introduction of Fermi-Dirac (Bose-Einstein)
functions for the quark (gluon) distributions \cite{bour96}. After many years
of research, we recently proposed \cite{BBS}, at the input scale
$Q_0^2 = 4 \mbox{GeV}^2$

\begin{eqnarray}
x u^{+}(x,Q^2_0) &=& {AX_{0u}^{+} x^b \over \exp[(x-X_{0u}^{+})/{\bar x}]
+1} + {\tilde A x^{\tilde b} \over \exp(x/{\bar x}) +1}\, , 
\label{eq5} \\
x \bar u^{-}(x,Q^2_0) &=& {\bar A (X_{0u}^{+})^{-1}x^{2b} \over 
\exp[(x+X_{0u}^{+})/{\bar x}]+1} + {\tilde A x^{\tilde b} \over 
\exp(x / {\bar x}) +1} \, , \label{eq6}\\
x G(x,Q^2_0) &=& {A_G x^{\tilde b +1} \over \exp(x /{\bar x}) - 1} \label{eq7}
\, ,
\end{eqnarray}
and similar expressions for the other light quarks ($u^{-},
d^{+}~\mbox{and}~ d^{-}$) and their antiparticles.
We assumed $\Delta G(x,Q^2_0) = 0$ and the strange quark distributions 
$s(x,Q^2_0)$ and $\Delta s(x,Q^2_0)$ are simply related \cite{BBS} to
$\bar q(x,Q^2_0)$ and $\Delta \bar q(x,Q^2_0)$, for $q = u,d$.
A peculiar aspect of this approach, is that it solves the problem of
desentangling the $q$ and $\bar q$ contribution through the relationship 
\cite{BBS}
\begin{equation}
  X_{0u}^{+} + X_{0 \bar u}^{-} = 0 \, ,
\label{eq11}
\end{equation}
and the corresponding one for the other light quarks and their antiparticles.
It allows to get the $\bar q(x)$ and $\Delta \bar q(x)$ distributions
from the ones for $q(x)$ and $\Delta q(x)$.
 
 By performing a next-to-leading order QCD evolution of these parton distributions, we
were able to obtain a good description of
a large set of very precise data on $F_2^{p,n}(x,Q^2), xF_3^{\nu N}(x,Q^2)$
and $g_1^{p, d, n}(x,Q^2)$, in correspondance with {\it eight} free 
parameters \cite{BBS}.
Therefore crucial tests will be provided by measuring flavor and spin 
asymmetries for antiquarks, for which we expect \cite{BBS} 
\begin{equation}
\Delta \bar u(x) > 0 > \Delta \bar d(x) \, ,
\label{eq12}
\end{equation}
\begin{equation}
\Delta \bar u(x) - \Delta \bar d(x) \simeq \bar d(x) - \bar u(x) > 0 \, .
\label{eq13}
\end{equation}
For illustration, we show in Fig.~1  
the predictions of the statistical approach with recent polarized
DIS results.

\begin{figure}[ht]
\centerline{\epsfxsize=3.1in\epsfbox{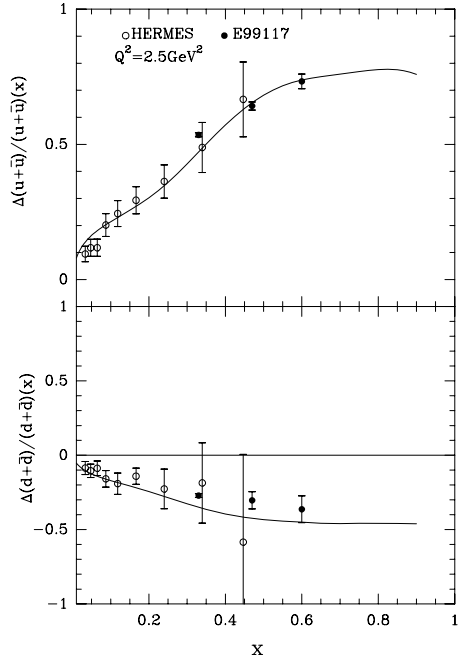}}   
\caption{Results for $\Delta(u + \bar u)/(u + \bar u)\,(x)$
and $\Delta(d + \bar d)/(d + \bar d)\,(x)$ from Refs.~[18,19],  
compared to the statistical model predictions.}
\label{jlab}
\end{figure}

The inequality $\bar d(x) - \bar u(x) > 0$ has the right sign to agree with the
defect in the GSR \cite{GOT}.
Let us make a few comments on the BSR. In the low $x$ region ( $x \leq 0.1$ ),
both $g_1^p$ and $g_1^n$ are expected to increase, as well as their difference.
The statistical prediction is close to the curve shown in Fig.~2, which has the
simple expression $0.18(x^{-0.5} - 1)$ implying 0.18 for the Bjorken integral.
The statistical model gives the value 0.176, in excellent agreement with the QCD prediction 
$0.182 \pm 0.005$ and with the world data $0.176 \pm 0.003 \pm 0.07$~\cite{E155}.
We also note that if Eq.~(\ref{eq13}) above is satisfied,
it means that the antiquark polarization contributes to the BSR. In the statistical
model this contribution is 0.022, which is not negligible. Finally this strong rise
in the low $x$ region, which was first noticed in Ref.~[21] is consistent with the
results from the resummation of double-logarithmic contributions \cite{EGT}.

\begin{figure}[ht]
\centerline{\epsfxsize=3.1in\epsfbox{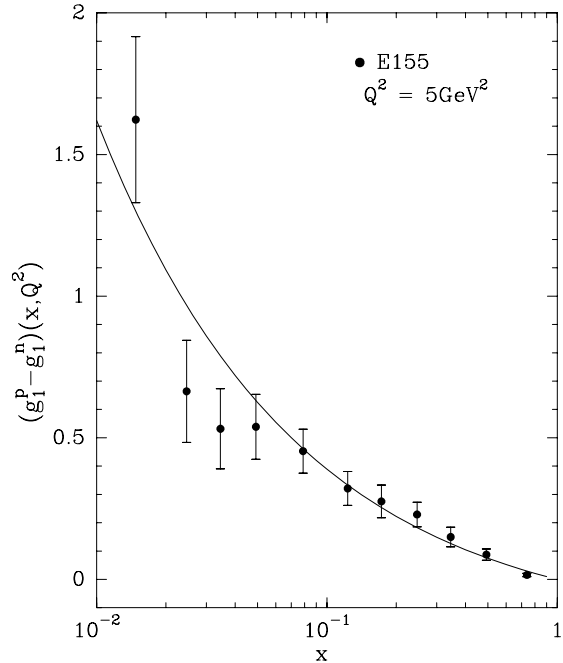}}   
\caption{The data on $g_1^{p-n}(x,Q^2)$ from Ref.~[20] compared to
the curve given by $0.18(x^{-0.5} - 1)$.}
\label{inter}
\end{figure}

\section{The generalized GDH sum rule revisited}
The generalized GDH sum rules
\cite{GERA} are just being tested experimentally for proton,
neutron and deuteron \cite{E143,JLAB,Hermes,JLAB2}.
The characteristic feature of the proton data is the strong dependence 
on the four-momentum transfer $Q^2$,
for $Q^2 <1 \mbox{GeV}^2$, with a zero crossing for $Q^2 \sim 200-250 \mbox{MeV}^2$,
which is in complete agreement
with our prediction \cite{ST93,ST95}, published almost ten years ago.
Our approach is making use of the relation to the BCSR
for the structure function $g_2$, whose elastic contribution
is the main source of a strong $Q^2$-dependence, while
the contribution of the other structure function, $g_T=g_1+g_2$ is smooth.
However, the recently published proton JeffLab data \cite{JLAB2} lie below
the prediction, displaying quite a similar shape.
Such a behaviour suggests, that
the reason for the discrepancy may be the oversimplified 
treatment of the QCD expressions at the boundary point $Q_0 \sim 1 \mbox{GeV}$,
defined in the smooth interpolation between large $Q^2$ and $Q^2=0$ and
which serve as an input for our model. For large $Q^2$ we 
took the asymptotic value for the GDH integral and we 
neglected all the calculable corrections, as well as the 
contribution of the $g_2$ structure function. 
This was quite natural and unnecessary ten years ago, since no data 
was available at that time.

In a recent work \cite{ST04}, we have filled up 
this gap by including the radiative (logarithmic) and power
QCD corrections. The procedure used to take these corrections into account
is explained in great details in Ref.~[29].
We found that the JeffLab data are quite sensitive to power corrections
and may be used for the extraction of the relevant phenomenological 
parameters. In Fig.~3 we display for $\Gamma^{p}_{1}(Q^2)$, the comparison
between the results of our previous work \cite{ST93,ST95} and the new
analysis which leads to a curve fairly close to the JeffLab data \cite{JLAB2}.
We show in Fig.~4 the same comparison for $\Gamma^{n}_{1}(Q^2)$
and 
we notice that the strong oscillation around 
$Q^2 =1\mbox{GeV}^2$, we had in the previous analysis, is no longer there.
These results can be easily transformed into predictions for the Bjorken
integral
$\Gamma^{p-n}_{1}(Q^2)$ and for the deuteron $\Gamma^{d}_{1}(Q^2)$ which turn
out to be in good agreement with preliminary JeffLab data \cite{DE,DO}, in the low
$Q^2$ region.

\begin{figure}[ht]
\centerline{\epsfxsize=3.5in\epsfbox{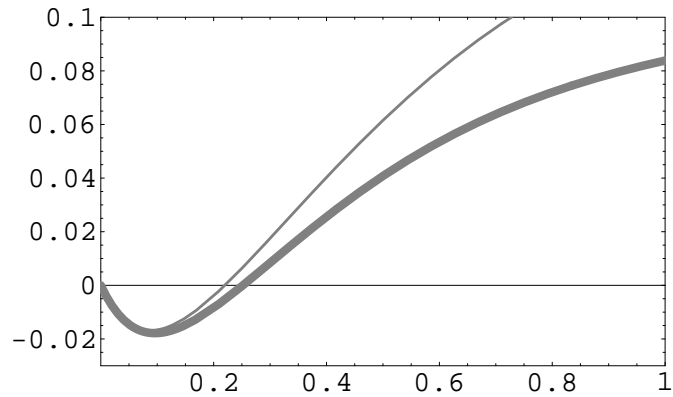}}
\caption{Our prediction for $\Gamma^{p}_{1}(Q^2)$. The thick line is the
new analysis 
{\protect \cite{ST04}}, to be compared with the thin line, which
represents our previous approach without corrections.}
\label{Fig3}
\end{figure}

\begin{figure}[ht]
\centerline{\epsfxsize=3.5in\epsfbox{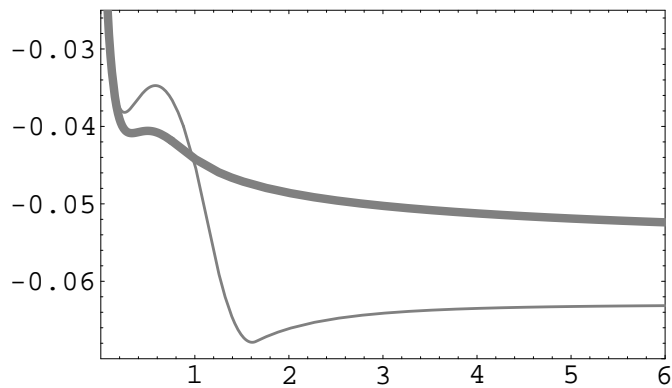}}
\caption{Our prediction for $\Gamma^n_1(Q^2)$. The thick line is the 
new analysis 
{\protect \cite{ST04}}, to be compared with the thin line, which
represents our previous approach without corrections.}
\label{Fig4}
\end{figure}

\newpage

\section*{Acknowledgments}
I would like to thank the organizers, in particular Sebastian Kuhn, for
the invitation and for giving me the opportunity to deliver this talk
at the very successful Symposium GDH2004.


\begin{thebibliography}{0}

\bibitem{BSSV} G. Bunce, N. Saito, J. Soffer and W. Vogelsang, {\it Ann. Rev.
Nucl. Part. Science} {\bf 50}, 525 (2000).

\bibitem{ADL} S. L. Adler, {\it  Phys. Rev.} {\bf 143}, 1144 (1966).

\bibitem{GRO} D. Gross and C. Llewellyn Smith, {\it Nucl. Phys.} 
{\bf B14}, 337 (1969).

\bibitem{ALLA} WA25 Collaboration, D. Allasia {\it et al.},
{\it Z. Phys} {\bf C28}, 321 (1985).

\bibitem{QUI} CCFR Collaboration, P. Z. Quintas {\it et al.},   
{\it Phys. Rev. Lett.} {\bf 71}, 1307 (1993); 
W. C. Leung {\it et al.}, {\it Phys. Lett.} {\bf B317}, 655 (1993);
J. H. Kim {\it et al.}, {\it Phys. Rev. Lett.} {\bf 81}, 3595 (1998).

\bibitem{GOT} K. Gottfried, {\it Phys. Rev. Lett.} {\bf 18}, 1154 (1967).


\bibitem{ARN} New Muon Collaboration, M. Arneodo {\it et al.}, 
{\it Phys. Rev.} {\bf D50}, R1 (1994) and references therein;
{\it Phys. Lett.} {\bf B364}, 107 (1995).


\bibitem{BJOR} J. D. Bjorken, {\it Phys. Rev.} {\bf 148} (1966),
1467 and {\it ibid} {\bf D1}, 1376 (1970).

\bibitem{ELL} J. Ellis and R. Jaffe, {\it Phys. Rev.} {\bf D9}, 1444 (1974);
{\bf D10} 1669 (E) (1974).

\bibitem{EMC} EMC Collaboration, J. Ashman {\it et al.}, 
{\it Phys. Lett.} {\bf B206}, 364 (1988).

\bibitem{BUR} H. Burkhardt and W. N. Cottingham, {\it Ann. Phys.}
{\bf 56}, 453 (1970).

\bibitem{JI} X. Ji, {\it Contribution to the Workshop on Deep Inelastic
Scattering and QCD}, Paris April 24-28, p. 435 (1995)
(Editors J. F. Laporte and Y. Sirois).

\bibitem{WAN} S. Wandzura and F. Wilczek, {\it Phys. Lett.}  {\bf B72}, 195 (1977).

\bibitem{GERA} S. B. Gerasimov, {\it Sov. J. Nucl. Phys.} {\bf 2}, 430
(1966); S. D. Drell and A. C. Hearn, {\it Phys. Rev. Lett.}  {\bf 16}, 908 (1966).


\bibitem{BBS} C. Bourrely, F. Buccella and J. Soffer, {\it Eur. Phys. J.} {\bf C23}, 487 (2002).\\
For a practical use of these PDF, see www.cpt.univ-mrs.fr/$\sim$bourrely \\
/research/bbs-dir/bbs.html.

\bibitem{BBS1} C. Bourrely, F. Buccella and J. Soffer, 
{\it  Mod. Phys. Lett.} {\bf A18}, 771 (2003).

\bibitem{bour96} C. Bourrely, F. Buccella, G. Miele, G. Migliore, J. Soffer and
V. Tibullo, {\it Z. Phys.} {\bf C62}, 431 (1994).

\bibitem{Hermes} Hermes Collaboration, K. Ackerstaff {\it et al.},
{\it Phys. Lett.}  {\bf B464}, 123 (1999).

\bibitem{jlab02} Jlab E-99-117 Collaboration, X. Zheng {\it et al.}, 
{\it Phys. Rev. Lett.}  {\bf 92}, 012004 (2004).

\bibitem{E155} SLAC E155 Collaboration, P. L. Anthony {\it et al.},
{\it Phys. Lett.}  {\bf B493}, 19 (2000).

\bibitem{ST} J. Soffer and O. Teryaev, {\it Phys. Rev.} {\bf D56}, 1549 (1997).

\bibitem{EGT} B.I. Ermolaev, M. Greco and S.I. Troyan, arXiv:hep-ph/0312029.

\bibitem{E143} E143 Collaboration, K. Abe et al., {\it Phys. Rev. Lett.} {\bf78}, 815 (1997).

\bibitem{JLAB}E94010 Collaboration, M. Amarian {\it et al.}, {\it Phys. Rev. Lett.} {\bf 89}, 
242301 (2002).

\bibitem{Hermes}HERMES Collaboration, A. Airapetian
{\it et al.}, 
{\it Phys. Lett.} {\bf B494}, 1 (2000);
 {\it Eur. Phys. J.} {\bf C26}, 527 (2003).

\bibitem{JLAB2} CLAS Collaboration, J. Yun {\it et al.}, {\it Phys. Rev.} {\bf C67}, 055204 (2003) and
R. Fatemi {\it et al.}, {\it Phys. Rev. Lett.} {\bf 91}, 222002 (2003).


\bibitem{ST93} J. Soffer and O. Teryaev, {\it Phys. Rev. Lett.} {\bf 70}, 3373 (1993).

\bibitem{ST95} J. Soffer and O. Teryaev, {\it Phys. Rev.} {\bf D51}, 25 (1995).

\bibitem{ST04} J. Soffer and O. Teryaev, preprint CPT-2004/P.017.

\bibitem{DE} A. Deur, these proceedings.

\bibitem{DO} G. Dodge, these proceedings.
 
\end{thebibliography}
\end{document}